\documentclass[A4paper,twoside]{article}
\usepackage{graphics}
\def\stacksymbols #1#2#3#4{\def\theguybelow{#2}
\def\verticalposition{\lower#3pt}
\def\spacingwithinsymbol{\baselineskip0pt\lineskip#4pt}
\mathrel{\mathpalette\intermediary#1}}
\def\intermediary#1#2{\verticalposition\vbox{\spacingwithinsymbol
\everycr={}\tabskip0pt
\halign{$\mathsurround0pt#1\hfil##\hfil$\crcr#2\crcr
\theguybelow\crcr}}}

\def\lapproxeq{\stacksymbols{<}{\sim}{4}{1}}

\title{Astrophysical Limits on Massive Dark Matter}
\author{G.Bertone$^{1,2}$, G.Sigl$^1$, J.Silk$^2$ \\
1. Institut d'Astrophysique, F-75014 Paris, France \\       
2. Department of Astrophysics, University of Oxford, NAPL 
\\ Keble Road, Oxford OX13RH, United Kingdom}
\date{}

\begin{document}

\maketitle

\label{firstpage}

\begin{abstract}
Annihilations of weakly interacting dark matter particles provide  an important signature for the possibility of indirect detection of dark matter in galaxy halos. These self-annihilations can be greatly enhanced in the vicinity of a massive black hole. We show that the massive black hole  present at the centre of our galaxy  accretes  dark matter particles, creating a region of very high particle density. Consequently the annihilation rate  is considerably increased, with a large number of $e^+e^-$ pairs being produced either directly or by successive decays of mesons. We evaluate  the synchrotron emission (and self-absorption) associated with  the propagation of these particles through the
galactic magnetic field, and are able to  constrain  the allowed values of
masses  and cross sections of dark matter particles. 
\end{abstract}

\section{Introduction}

There is convincing evidence for the existence of an unseen non-baryonic
component in the energy-density of the universe. The most promising dark
matter candidates appear to be weakly interacting massive particles
(WIMPs) and in particular the so-called neutralinos, arising in
supersymmetric scenarios as well as much heavier particles (WIMPzillas)
which could have been produced non-thermally in the early universe (for a
review of particle candidates for dark matter see e.g. Ellis (1998)). 

The annihilations of these X-particles would generate quarks, leptons, gauge
and Higgs bosons and gluons. Consequently $e^+e^-$ pairs would be produced
either directly or by successive decays of mesons, and they are expected
to lose their energy through 
synchrotron radiation as they propagate in the galactic magnetic field.
This radiation is expected to be greatly enhanced in the
proximity of the galactic centre, where the existence of a massive black
hole creates a region of very high dark matter particle density and
consequently a great increase in the annihilation rate and synchrotron
radiation.

We discuss in section 2 the distribution of dark matter particles in our
galaxy and in particular around the central black hole, following Gondolo
\& Silk (1999) (from now on, Paper I). In section 3 we evaluate the
annihilation $e^+e^-$ spectrum and their synchrotron emission. Finally in
section 4 we discuss our results and determine the regions of the {\it
mass-annihilation cross section} plane which give predictions compatible
with experimental data.

\section{Dark matter distribution around the galactic centre}

There is strong evidence for the existence of a massive compact object
lying within the inner $0.015 \rm pc$ of the galactic centre (see
Yusef-Zadeh, Melia  \& Wardle (2000) and references therein). This object
is a compelling candidate for a massive black hole, with  mass $M=(2.6 \pm 0.2)\times10^6 M_{\odot}$.
The galactic halo density profile is modified in the neighborood of the
galactic centre from the adiabatic process of accretion towards the
central black hole. If we consider an initially power-law type profile of
index $\gamma$, 
similar to  those predicted by high resolution N-body simulations
(Navarro, Frenk \& White 1997; Ghigna et al. 2000), the corresponding dark matter profile
after accretion is, from paper I
\begin{equation}
\rho' = \left[ \alpha_{\gamma} \left( \frac{M}{\rho_D
D^3}\right)^{3-\gamma} \right]^{\gamma_{sp}-\gamma} \rho_D \; g(r) \left(
\frac{D}{r} \right)^{\gamma_{sp}}
\label{modaccre}
\end{equation}
where $\gamma_{sp}=(9-2\gamma)/(4-\gamma)$, $D$ is the solar distance from
the galactic centre and $\rho_D=0.24 GeV/c^2/cm^3$ is the density in the
solar neighbourhood. The factors $\alpha_\gamma$ and $g_\gamma(r)$ cannot
be determined analytically (for approximate expressions and numerical
values see paper I). The expression~(\ref{modaccre}) is only valid in a
central region of size $R_{sp}=\alpha_\gamma D (M/\rho_D
D^3)^{1/(3-\gamma)}$ where the central black hole dominates the
gravitational potential.

If we take into account the annihilation of dark matter particles, the
density cannot grow to arbitrarily high values, the maximal density being
fixed by the value 
\begin{equation}
\rho_{core}=m/\sigma v t_{BH}
\label{roco}
\end{equation}
where $t_{BH}\approx 10^{10}years$ is the age of the central black hole.
The final profile, resulting from the adiabatic accretion of annihilating
dark matter on a massive black hole is
\begin{equation}
\rho_{dm}(r)= \frac{\rho'(r) \rho_{core}}{\rho'(r)+\rho_{core}}
\end{equation}
following a power law for large values of r, and with a flat core of
density $\rho_{core}$ and  dimension 
\begin{equation}
R_{core}=R_{sp} \left( \frac{\rho(R_{sp})}{ \rho_{core}} \right)
^{(1/\gamma_{sp})}
\end{equation}

\section{Constraints from Synchrotron Emission}
Among the products of annihilation of dark matter particles, there will be
electrons and positrons, which are expected to produce synchrotron
radiation in the magnetic field around the galactic centre. The $e^+e^-$
component produced by the hadronic jets 
has been computed using the MLLA approximation\footnote{For details see
Dokshitzer et al. (1991), Ellis, Stirling, \& Webber (1996), Khoze \& Ochs
(1997). For applications to ultra-high energy cosmic rays see
Bhattacharjee \& Sigl (2000).}.

The galactic magnetic field can be described using the 'equipartition
assumption', where the magnetic, kinetic and gravitational energy of the
matter accreting on the central black hole are in approximate
equipartition (see Melia (1992)). In this case the magnetic field can be
expressed as
\begin{equation}
B(r) = 1\mu G \left( \frac{r}{pc} \right)^{-5/4}
\label{mag}
\end{equation}

\begin{figure}[t]
\resizebox{11.5cm}{!}{\includegraphics{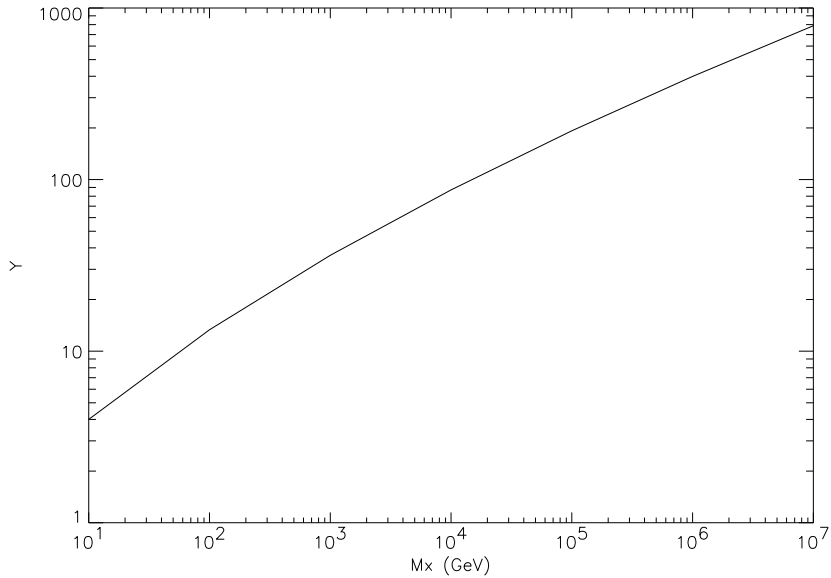}}
\caption[Values of $Y_e$ as a function of particle mass]{Values of $Y_e$
as a function of particle mass.}
\label{Y}
\end{figure}

\paragraph{Energy-loss length scale. } As we shall see, most of the
annihilations happen at very small distances from the centre, typically at
$\approx min(R_{core}, 10 Rs)$, i.e. in a region with magnetic fields of
the order of $> 1G$. Under these conditions, comparable to the size of the
region where most of the annihilations occur, the electrons lose their
energy almost in place. To see this, consider the critical synchrotron
frequency $\nu_c(E)$ i.e. the frequency around which the synchrotron
emission of an electron of energy E, in a magnetic field of strength B,
peaks, namely
\begin{equation}
\nu_c(E) = \frac{3}{4 \pi} \frac{eB}{m_e c} \left(\frac{E}{m_e c^2}
\right)^2\,.
\end{equation}
Inverting this relation, we find the energy of the electrons which give
the maximum contribution at that frequency,
\begin{equation}
E_m(\nu) = \left( \frac{4 \pi}{3} \frac{m_e^3 c^5}{e} \frac{\nu}{B}
\right)^{1/2} = 0.25 \left( \frac{\nu}{\mbox{MHz}} \right)^{1/2}  \left(
\frac{r}{\mbox{pc}} \right)^{5/8} \mbox{GeV}
\label{em}
\end{equation}
\begin{figure}[t]
\centering
\resizebox{11.5cm}{!}{\includegraphics{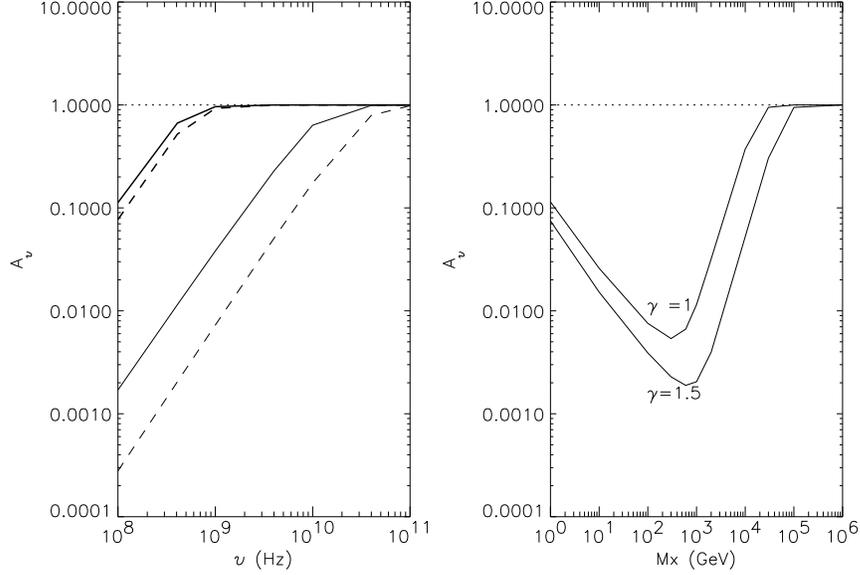}}
\caption{Left panel: $A_{\nu}$ as a function of frequency for $m_X=1$TeV.
The two upper curves correspond to the cross section
$\sigma v\approx 10^{-28}/m_X^2(\rm GeV)\,\rm cm^3s^{-1},$
% $\sigma v=1/m_X^2$,
close to the unitarity limit; the two lower curves correspond to
$\sigma v\approx 10^{-38}/m_X^2(\rm GeV)\,\rm cm^3s^{-1},$
%$\sigma v=10^{-10}/m_X^2$, 
a cross section more typical for wimps.
Results for two values of the density profile are shown in each case:
$\gamma=1$ (solid curves) and $\gamma=1.5$  (dashed curves).
Right panel: $A_{\nu}$ as a function of
the particle mass for $\nu$=408MHz, $\sigma v=10^{-10}/m_X^2$ 
(in physical units) and two
values of $\gamma$. \label{absnu}}
\end{figure}
The typical synchrotron loss length for an energy corresponding to
$E_m(\nu)$ can be expressed  as a function of the distance $r$ from the
central black hole for the magnetic field profile in Eq.~(\ref{mag}):
\begin{eqnarray}
l_e(E_m(\nu))=  \left( \frac{27}{16 \pi} \right)^{1/2}  \left( \frac{m_e^5
c^{11}}{e^7} \right)^{1/2} \frac{1}{\nu^{1/2}} \frac{1}{B^{3/2}} &
\nonumber \\ = 1.025 \cdot 10^{10}\left( \frac{\nu}{\mbox{MHz}}
\right)^{-1/2} \left( \frac{r}{\mbox{pc}} \right)^{15/8} \mbox{pc}&
\end{eqnarray}
For a frequency of 408 MHz, which produces, as we shall see, the most
stringent upper bound on the Galactic Centre emission, we get 
\begin{equation}
l_e(E_m(\nu))= 5.074 \cdot 10^{8}  \left( \frac{r}{\mbox{pc}}
\right)^{15/8} \mbox{pc}
\end{equation}
The diffusion length $D=D (B(r),E)$, which can be approximated as a third
of the radius of gyration of the electron, is given by
\begin{equation}
D(E_m(\nu))= \frac{E}{3eB} \approx 8.991 \cdot 10^{-8} \left(
\frac{\nu_c}{\mbox{MHz}} \right)^{1/2} \left( \frac{r}{\mbox{pc}}
\right)^{15/8}\mbox{pc}
\end{equation}
Since $l_e(E_m(\nu))>>D(E_m(\nu))$ for all practically relevant
parameters, an electron will diffuse a distance $\sqrt{l_e D}$ before it
loses most of its energy. Numerically,
\begin{equation}
\frac{\sqrt{l_e D}}{r} \approx 3.035  \cdot 10^{1}  \left(
\frac{r}{\mbox{pc}} \right)^{7/8} 
\end{equation}
or, expressing the  distance as a function of the Schwarzschild radius
\begin{equation}
\frac{\sqrt{l_e D}}{r} \approx 7.059 \cdot 10^{-5}  \left( \frac{r}{R_s}
\right)^{7/8} 
\end{equation}
We can thus assume that the electrons lose their energy practically in
place.
\begin{figure}[t]
\resizebox{11.5cm}{!}{\includegraphics{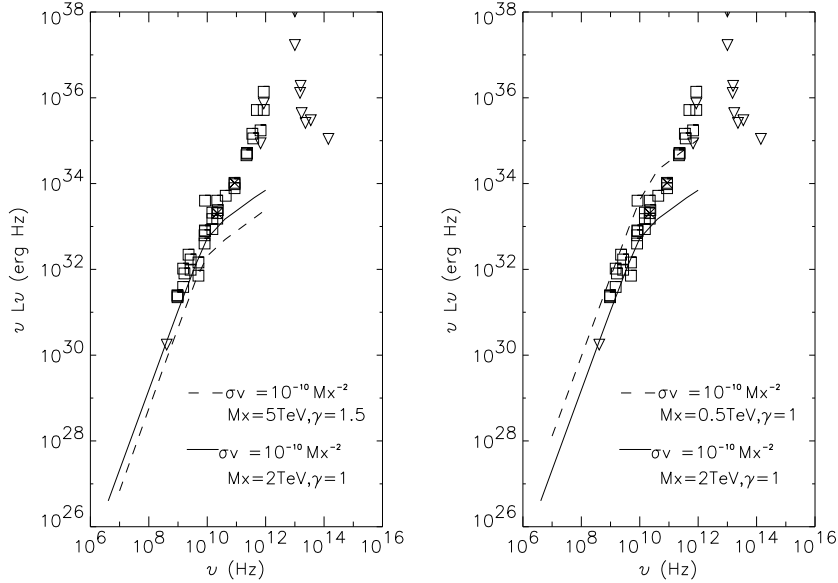}}
\caption[Comparison of Sgr A* observed spectrum with expected
fluxes]{Comparison of Sgr A* (see Narayan et al. (1998)) observed spectrum
with expected fluxes. The values of particle mass and cross section were
choosen to fit the experimental data normalisation.}
\label{predi1}
\end{figure}

\paragraph{Electron Production Spectrum. } To compute the synchrotron
luminosity produced by the propagation of $e^{\pm}$ in the galactic
magnetic field, we need to evaluate their energy distribution in the
magnetic field which is given by (see Gondolo (2000))
\begin{equation}
 \frac{dn}{dE} = \frac{\Gamma Y_e(>E)}{P(E)} f_e(r)
\label{dn}
\end{equation}
where $\Gamma$ is the annihilation rate
\begin{equation}
\Gamma = \frac{\sigma v}{m_x^2} \int_0^{\infty} \rho_{sp}^2 4 \pi r^2 \;\;
dr\,.
\end{equation}
The function $f_e(r)$ is given by
 \begin{equation}
f_e(r)= \frac{\rho_{sp}^2}{  \int_0^{\infty} \rho_{sp}^2 4 \pi r^2 \;\;
dr}
\end{equation}
and 
 \begin{equation}
P(E)= \frac{2 e^4B^2 E^2}{3 m_e^4 c^7}
\end{equation}
is the total synchrotron power spectrum.
Note that the general expression for $f_e(r)$ would have to take into
account spatial redistribution by diffusion (see e.g. Gondolo (2000)),
which we demonstrated to be negligible in our model.

The quantity $Y_e(>E)$ is the number of $e^+e^-$ with energy above E
produced per annihilation, which depends on the annihilation modes.
Eq.~(\ref{em}) shows that for the frequencies we are interested in,
$E_m(\nu)<<m_X$, and thus the energy dependence of
 $Y_e(>E)$ can be neglected. 
We will estimate $Y_e(>E)$ by the number of charged particles produced in
quark fragmentation (see footnote); in figure \ref{Y} we show the values
of Y as a function of the particle mass.

\paragraph{Synchrotron Luminosity. } For each electron the total power
radiated in the frequency interval between $\nu$ and $\nu + d\nu$ is given
by
\begin{eqnarray}
P(\nu,E) = \frac{\sqrt{3} e^3}{m_e c^2} B(r) \frac{\nu}{\nu_c(E)}
\int_0^{\infty} K_{5/3}(y) dy & \nonumber \\ = \frac{\sqrt{3} e^3}{m_e
c^2} B(r) F\left( \frac{\nu}{\nu_c(E)} \right)&
\end{eqnarray}
where we introduced
\begin{equation}
F\left( \frac{\nu}{\nu_c(E)} \right)=  \frac{\nu}{\nu_c(E)}
\int_0^{\infty} K_{5/3}(y) dy
\end{equation}
Integrating this formula we obtain the total synchrotron luminosity
\begin{equation}
L_{\nu} = \int_0^{\infty} dr \; \;  4 \pi r^2  \int_{m_e}^{m_x} dE \;\;
\frac{dn_e}{dE} P(\nu,E)
\label{lum}
\end{equation}
which by substitution becomes
\begin{equation}
L_{\nu} =\frac{\sqrt{3} e^3 \Gamma}{m_e c^2} \int_0^{\infty} dr 4 \pi r^2
f_e(r) B(r)  \int_{m_e}^{m_x} dE \frac{Y_e(>E)}{P(E)} F\left(
\frac{\nu}{\nu_c(E)} \right)
\end{equation}

It is possible to simplify this formula by introducing the following
approximation for the function $F\left( \frac{\nu}{\nu_c(E)} \right)$ (see
Rybicki  \&  Lightman (1979))
\begin{equation}
F\left( \frac{\nu}{\nu_c(E)} \right) \approx \delta(\nu / \nu_c(E) -
0.29)\,.
\label{approx}
\end{equation}
The evaluation of the integral then gives
\begin{equation}
L_{\nu} \approx \frac{9}{8} \left(\frac{1}{0.29 \pi} \frac{ m_e^3 c^5}{e}
\right)^{1/2} \frac{\Gamma Y_e(>E)}{\sqrt{\nu}} \; I
\end{equation}
where 
\begin{equation}
I = \int_0^{\infty} dr \;\; 4 \pi r^2 f_e(r) B^{-1/2}(r) 
\end{equation}

\paragraph{Synchrotron Self-Absorption. } The synchrotron self-absorption
coefficent is by definition (see Rybicki  \&  Lightman (1979))
\begin{equation}
A_{\nu}= \frac{1}{a_{\nu}} \int_{0}^\infty (1-e^{-\tau(b)}) 2 \pi b \;\;
db
\label{anu}
\end{equation}
where $\tau(b)$ is the optical depth as a function of the cylindrical
coordinate b
\begin{equation}
\tau(b)= a_{\nu} \int_{d(b)}^{\infty} f_e(b,z) \;\;dz
\label{la}
\end{equation}
and the coefficent $a_{\nu}$ is given by
\begin{equation}
 a_{\nu} =\frac{e^3 \Gamma B(r)}{9 m_e \nu^2} \int_{m_e}^{m} E^2 \frac
{d}{dE} \left[ \frac{Y_e(>E)}{E^2 P(E)} \right] F\left( \frac{\nu}{\nu_c}
\right) \;\; dE
\end{equation}

The final luminosity is obtained by multiplying Eq.~(\ref{lum}) with
$A_{\nu}$ given by Eq.~(\ref{anu}). It is evident that in the limit of
small optical depths the coefficent $A_{\nu} \rightarrow 1$, as can be
seen by expanding the exponential.

Furthermore the lower limit of integration of expression \ref{la} is 
\begin{equation}
d(b)= 0 \;\;\;\; for \;\;\;\;  b^2+z^2>(4Rs)^2 
\end{equation}
\[ d(b)= \sqrt{(4Rs)^2-b^2} \;\;\;\;\; elsewhere \]

Using the approximation introduced in Eq.~(\ref{approx}) we find for
$a_{\nu}$ the following expression
\begin{equation}
a_{\nu} =\frac{\Gamma Y}{4 \pi} \frac{c^2}{\nu^3},
\end{equation}
which can in turn be used to evaluate $\tau(b)$ in Eq.~(\ref{la}) and
$A_{\nu}$ in Eq.~(\ref{anu}).
\begin{figure}[t]
\resizebox{11.5cm}{!}{\includegraphics{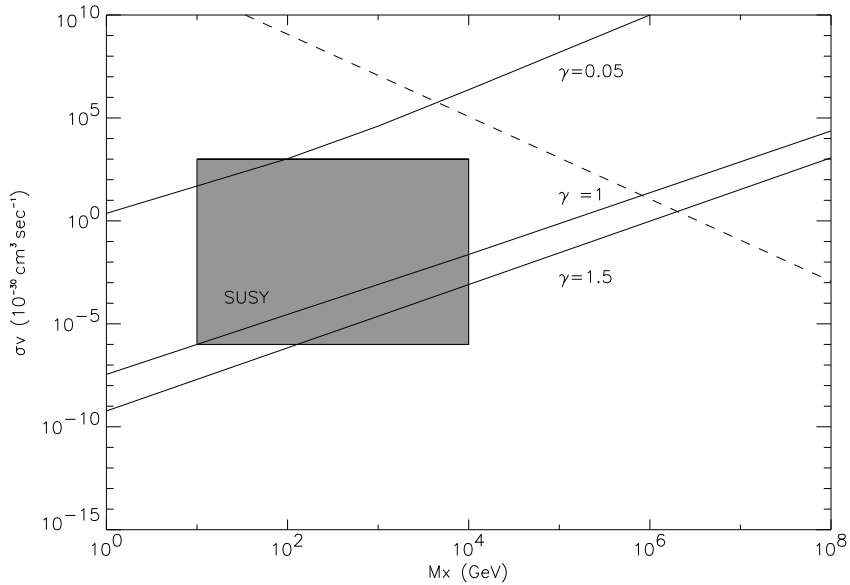}}
\caption[Exclusion plot based on upper limit at 408MHz]{Exclusion plot
based on the comparison between predicted flux and radio observations of
the galactic centre.The 3 solid curves indicate, for different values of
the density profile power law index, the lower edge of the excluded
regions. The dashed line shows, for comparison, the unitarity bound,
$\sigma v\simeq1/m_X^2$. The shaded region is the portion  of the
parameter space occupied by cosmologically interesting neutralinos (i.e.
those leading to 
$0.025<\Omega_X h^2<1$; see, e.g. Bergstrom, Ullio \& Buckley (1997)).}
\label{excludo}
\end{figure}

  Note that the synchrotron
loss  time-scale  is proportional to $B^{-2} \propto r^{5/2}.$
We compare this with  the annihilation time, which  determines where the
inner density drops drastically:
$t_{ann} \propto r^{(9-2\gamma)/(4-\gamma) + 1/2}.$ In other words, the
annihilation time goes to zero more rapidly than 
the synchrotron loss time as $r$ goes to 0. Hence synchrotron losses are
important throughout the annihilation region.

Extension to the  wimpzilla 
mass range  introduces far more uncertain physics.
The wimpzilla cross sections strongly depend on new physics
  beyond the electroweak scale. Similarly to the case of electroweak
  gauge bosons, one can expect a general scaling $\sigma v\sim
  \alpha^2/m_X^2$ for the self-annihilation cross section where
  $\alpha \lapproxeq 1$ is some dimensionless gauge coupling. In four
  dimensional field theory, the conservation of probability (unitarity)
  roughly corresponds to this scaling for
  $\alpha\to1$ (see, e.g., Weinberg (1995)).
  In order to retain generality, we will therefore consider
  Wimpzilla annihilation cross sections in the range $\sigma v
  \lapproxeq 1/m_X^2$. We note, however, that beyond four dimensional
  field theory, such as in theories with extra dimensions and
  in string theory, larger cross sections may be possible.

We first evaluate the self-absorption coefficent for selected values of
the mass, as a function of frequency. 
The result, shown in fig.\ref{absnu}, is a coefficent which grows from
very low values (showing that absorption is important at 408 MHz) and then
reaches the value 1, around a frequency which is strongly dependent on the
cross section and the mass $m_X$, 
but not so much on the profile power law index $\gamma$. On the left the
above coefficent is evaluated for two different values of the cross
section, the first one corresponding to the maximum possible value (the so
called unitary bound, see Griest \& Kamionkowski (1990))
 and the second one corresponding to typical cross sections one can find
in supersymmetric scenarios (see, e.g., Bergstrom, Ullio \& Buckley
(1997)).

The right part of fig.\ref{absnu} shows  the self-absorption coefficent at
the fixed frequency of 408MHz as a function of the neutralino mass. The
behaviour shown is qualitatively the same for any value of the
cross-section and for different $\gamma$. 
In figure \ref{predi1} we compare the predicted spectrum with the
observations; we choose a  set of parameters $m_X$, $\gamma$ and $\sigma
v$ in order to reproduce the observed normalisation. It is remarkable that
in that way one can reproduce the observe
d spectrum over a significant range of frequencies.
The set of dark matter parameters for which the fluxes predicted in our
model are consistent with observation is shown in the exclusion plot of
fig. \ref{excludo}. The boundaries of the excluded range represent the
parameter values for which the observed 
flux is explained by the dark matter scenario discussed here.

\section{Conclusions and perspectives}
\label{concludo}
We have  shown that present data on the emission from Sgr A*  are
compatible with quite a wide set of dark matter parameters.
The evaluation of synchrotron self-absorption has enabled us to reach an
alternative conclusion from an earlier claim of incompatibility of cuspy
halos with the existence of annihilating wimp dark matter.
Our final result is that the experimental data on Sgr A* spectrum at radio
wavelengths could be explained by synchrotron emission of electrons
produced in the annihilation of relatively  massive dark matter particles,
extending from TeV masses to $m_X > 10^8 \rm GeV$. The former is relevant to recent studies  of coannihilation (Boehm,
Djouadi \& Drees (2000)), that suggest that WIMPs with  $\Omega_X h^2=0.2$
can extend up to several TeV, and the latter is relevant for particles
(wimpzillas) that are produced non-thermally in the primordial universe. 

We have found that with the current data situation, the synchrotron
emission tends to give somewhat sharper constraints on masses and cross
sections than the gamma-ray fluxes (cf. Baltz et al. (2000)). This
situation could be reversed by more sensitive  gamma-ray observations
anticipated from  upcoming experiments.
However the synchrotron predictions
are uncertain because of our relative
ignorance about the magnetic field strength near the central black hole,
and gamma ray fluxes are subject to similarly uncertain amounts
 of self-absorption.The ANTARES neutrino experiment will eventually be
able to set a relatively model-independent
limit on the annihilation flux from the Galactic Centre.

\label{lastpage}
\end{document}